\documentclass[conference]{IEEEtran}
\IEEEoverridecommandlockouts
\usepackage{listings}
\usepackage{tabularx}

\lstdefinestyle{mystyle}{
    basicstyle=\ttfamily\footnotesize,
    breakatwhitespace=false,         
    breaklines=true,                 
    captionpos=b,                    
    keepspaces=true,                 
    showspaces=false,                
    showstringspaces=false,
    showtabs=false,                  
    tabsize=2
}

\usepackage{cite}
\usepackage{bytefield}
\usepackage{amsmath,amssymb,amsfonts}
\usepackage{algorithmic}
\usepackage{graphicx}
\usepackage{textcomp}
\usepackage{xcolor}
\usepackage{etoolbox}
\usepackage{enumitem}
\usepackage{tikz}
\usepackage[hidelinks]{hyperref}
\def\BibTeX{{\rm B\kern-.05em{\sc i\kern-.025em b}\kern-.08em
T\kern-.1667em\lower.7ex\hbox{E}\kern-.125emX}}
\begin{document}

\title{Secure and Dynamic Publish/Subscribe: LCMsec}

\author{\IEEEauthorblockN{Moritz Jasper,
Stefan Köpsell}
\IEEEauthorblockA{\textit{Barkhausen Institut gGmbH}, Würzburger Straße 46, Dresden, Germany }
\textit{\{moritz.jasper, stefan.koepsell\}@barkhauseninstitut.org}
}

\maketitle

\begin{abstract}
    We propose LCMsec, a brokerless, decentralised Publish/Subscribe protocol.
    It aims to provide low-latency and high-throughput message-passing for IoT
    and automotive applications while providing much-needed security
    functionalities to combat emerging cyber-attacks in that domain. LCMsec is
    an extension for the Lightweight Communications and Marshalling (LCM)
    protocol. We extend this protocol by providing not only authenticated
    encryption of the messages in transit, but also a group discovery protocol
    inspired by the Raft consensus protocol. The Dutta-Barua group key
    agreement is used to agree upon a shared symmetric key among subscribers
    and publishers on a topic. By using a shared group key, we reduce the key
    agreement overhead and the number of message authentication codes (MACs)
    per message compared to existing proposals for secure brokerless
    Publish/Subscribe protocols, which establish a symmetric key between each
    publisher and subscriber and append multiple MACs to each message.
\end{abstract}


\section{Introduction}

Publish/Subscribe architectures \cite{eugsterManyFacesPublish2003} are widespread and an important building block
for Internet of Things (IoT), automotive and cloud applications. They can
improve scalability and flexibility of communication infrastructures by
decreasing dependencies between components, since entities  in such a system
need not know about one another. They additionally support dynamic
communication patterns in which publishers and subscribers can be added and
removed without affecting the rest of the system. Some Publish/Subscribe
protocols like the Lightweight Communication and Marshalling protocol (LCM)
\cite{huangLCMLightweightCommunications2010} are brokerless, which offers
advantages in terms of latency and throughput in some situations, removes a
central point of failure (the broker) and reduces the administrative overhead.

However, LCM fails to offer convenient and fast possibilities of securing it.
There exists no easy way to achieve security by leveraging existing
transport-layer encryption mechanisms due to the multicast-based communication
topology that is used in LCM: achieving security in the multicast case is
generally a much harder problem than in the unicast case
\cite{canettiMulticastSecurityTaxonomy1999}. Thus, LCM, even when used in an
isolated network, not only violates the emerging zero-trust paradigm but also
the need-to-know principle: messages are simply routed to all other users of
the system, even those that have not subscribed to the particular topic.

Nevertheless, the brokerless Publish/Subscribe communication topology offers
the distinct advantages in terms of latency, throughput and simplicity mentioned above. The purpose
of this work is therefore to provide an extension to LCM, which preserves the
benefits in performance and ease of usability. Furthermore, it ensures confidentiality,
integrity and authenticity for the messages in transit.

An overview and evaluation of the existing security solutions in the
Publish/Subscribe space is discussed in Section \ref{sec: Work}. In Section \ref{sec:Description
of LCM}, we discuss the LCM protocol in detail since it forms the basis for this work. After defining an attacker model and security goals in Section \ref{sec:Attacker Model and Security
Goals}, we present the proposed LCMsec protocol in Section \ref{sec: protocol_description},
which contains two phases: firstly, the scheme used to secure messages based
on shared keying material, secondly, the scheme used to agree on that keying
material. Finally, we evaluate the performance of the proposed protocol in Section
\ref{sec:Implementation}.

\section{Related Work}%
\label{sec: Work}

\subsection{Publish/Subscribe Systems}%
\label{ssub:Publish/Subscribe Systems}

Typically, a distinction is made between topic-based and content-based
Publish/Subscribe systems \cite{eugsterManyFacesPublish2003}. In a topic-based
system, subscribers can subscribe to one or multiple topics. Messages in such a
system are associated with a specific topic, and receivers will only receive
messages on topics they are interested in. In a content-based system,
subscribers can instead express constraints on the contents of messages
directly.

Furthermore, Publish/Subscribe systems usually adopt either a brokered or
brokerless architecture.
Brokered systems like the widely used Message Queue Telemetry Transport (MQTT)
\cite{MessageQueuingTelemetry} use a central message broker to transmit
messages between the publishers and subscribers. This allows fine-grained control
over message distribution since brokers can route messages based on the
constraints of the subscribers (whether they are content- or topic-based).

Brokerless Publish/Subscribe systems distribute messages directly from
publishers to subscribers in a peer-to-peer fashion, which can improve latency
and throughput characteristics while reducing the amount of configuration that
is required to deploy entities. Additionally, the decentralised nature of such
systems does not depend on a single point of failure. Examples for such systems
include the Data Distribution Service (DDS)
\cite{pardo-castelloteOMGDataDistributionService2003} and LCM, both of which
can use UDP over IP multicast \cite{deeringHostExtensionsIP1989} for message
delivery to achieve high-throughput and low-latency in scalable systems.

\subsection{Security in Publish/Subscribe Systems}%
\label{ssub:Security in Publish/Subscribe Systems}

Most work that proposes security solutions for Publish/Subscribe systems
focuses on brokered Publish/Subscribe architectures. For instance, Onica et
al. \cite{onicaScalableDependablePrivacyPreserving2016} stated a list of
requirements for privacy-preserving Publish/Subscribe systems, but consider
only systems which use a broker.
Bernard et al. \cite{bernardFrameworkSecurePrivate2010} proposed a general,
conceptual framework for peer-to-peer data exchange that can also be used
with existing Publish/Subscribe systems, although brokers are used in this
scenario. Malina et al. \cite{malinaSecurePublishSubscribe2019} proposed a
security framework for MQTT which uses brokers. Ion et al.
\cite{ionSupportingPublicationSubscription2010} and Hamad et al.
\cite{hamadSPPSSecurePolicybased2021} described systems in which brokers are employed but
not trusted. Similarly, Dahlmanns et al. propose ENTRUST
\cite{dahlmannsTransparentEndtoEndSecurity2021}, achieving end-to-end security
over any existing brokered Publish/Subscribe system without trusting those brokers.

ZeroMQ \cite{imatixcorporationZeroMQBrokerVs}
can be used to implement brokerless Publish/Subscribe messaging, however, there
are no security extensions for it with support for this use-case. CurveZMQ
\cite{imatixcorporationCurveZMQProtocolSecure2023}, while similar in name, is
quite different and does not actually provide security for Publish/Subscribe
systems, but end-to-end security between client and server. While CurveZMQ can
be used to secure Publish/Subscribe by being embedded in the transport layer,
this is only possible when client and server are only one hop apart.

The Data Distribution Service (DDS), however, is quite comparable to LCM with
regard to their respective use-cases. DDS supports the brokerless
Publish/Subscribe paradigm in a peer-to-peer fashion, that is without using a
message broker, however, it works slightly differently to LCM. Instead of
simply broadcasting messages to a preconfigured multicast group, DDS features a
discovery protocol that allows publishers to discover the set of appropriate
subscribers. Subsequently, messages are routed only to these subscribers.

DDS also features a security extension
\cite{objectmanagementgroupDDSSecuritySpecification2018} that provides
authenticated encryption on a per-message basis. However, a handshake and key
agreement is performed separately between each publisher and subscriber to a topic (as discovered
by the discovery protocol)\cite{kimSecurityPerformanceConsiderations2018}.
This may lead to scalability issues during the discovery phase in the case
of large numbers of publishers or subscribers to the same topic. A high amount
of flexibility and many ways to configure the DDS middleware can lead to
misconfiguration, a problem which is also mentioned in
\cite{kimSecurityPerformanceConsiderations2018}. Additionally, there are
scalability issues at runtime. Authentication of messages is achieved by using
a separate Message Authentication Code for each receiver
\cite{objectmanagementgroupDDSSecuritySpecification2018}
which, in the case of many subscribers, leads either
to large overhead for each message or separate messages for each receiver,
moving away from the multicast paradigm.

These scalability issues are quite inherent to the problem of authenticating
messages in a multicast setting in which digital signatures are not desired due
to their poor performance characteristics. While a number of theoretical
solutions are discussed in literature
\cite{canettiMulticastSecurityTaxonomy1999}, we bypass this problem entirely.
By defining a trusted group of legitimate publishers and subscribers that share
a common symmetric, ephemeral key, we propose a protocol in which an authentic
message is understood to be a message originating from any member of this
group, not necessarily a specific one. In order to generate this shared key
while avoiding a scenario in which a total of $N\cdot M$ expensive key agreements
need to be carried out (in the case of $N$ publishers and $M$ subscribers), we
use the Dutta-Barua group key agreement (DBGKA)
\cite{duttaProvablySecureConstant2008}, an authenticated group key agreement
protocol that supports dynamic joining and leaving of users. Furthermore, we
implement a discovery protocol, inspired by the Raft consensus algorithm
\cite{ongaroSearchUnderstandableConsensus2014}, that forms consensus about the
state of the trusted group in order to drive the DBGKA protocol.

\section{Description of LCM}
\label{sec:Description of LCM}

Lightweight Communications and Marshalling \cite{huangLCMLightweightCommunications2010} is a brokerless, topic-based
Publish/Subscribe protocol designed for real-time systems that require
high-throughput and low-latency. Message types can be defined in the LCM type
specification language, which is a language-neutral and platform-neutral
specification language for structured data. From this specification language,
language-specific bindings for binary serialisation and encoding are generated,
while maintaining interoperability.

The binary-encoded LCM messages are then sent via multicast groups, which are
identified by the multicast IP-address and port on which they are transmitted.
Each group comprises multiple topics, which in LCM are called channels,
identified by a channelname string. Messages are transmitted using UDP and
routed via IP-multicast to all other nodes within the multicast group. A node
can subsequently subscribe to a channel within that group by simply dropping
all messages except those that match the\textit{ channelname}. Since the same \textit{channelname} might be used in multiple\textit{ multicastgroups} at the same time, we can uniquely identify a only by the
\textit{combination} of \textit{multicastgroup} and \textit{channelname}. We will therefore
define \textit{LCMDomain=(multicastgroup, channelname)}.

\begin{figure}[!htb]
\begin{center}
    \includegraphics[scale=0.7]{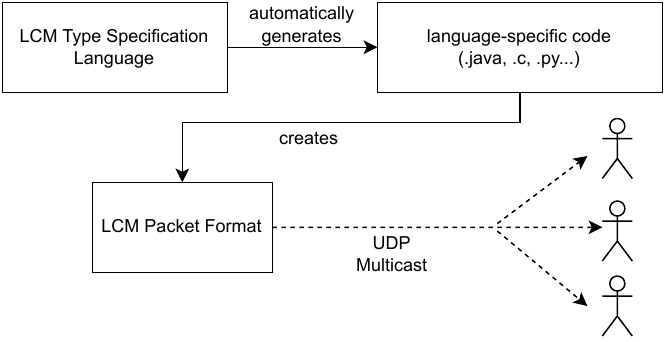}
\end{center}
\caption{High-level illustration of LCM}
\label{fig:high_level_lcm}
\end{figure}

\begin{figure}[!htb]
    \begin{center}
        \includegraphics[scale=0.8]{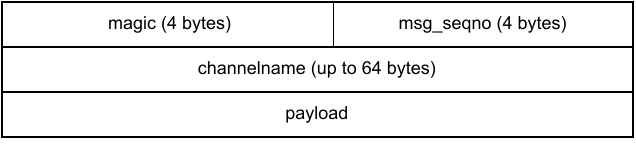}
    \end{center}
    \caption{LCM packet format}
    \label{fig:vanilla_lcm_header}
\end{figure}

The LCM packet format, as depicted in Figure \ref{fig:vanilla_lcm_header}, consists of a 4 byte magic number to identify the LCM protocol, a sequence number which is incremented by each sender separately, and a zero-terminated, ASCII-encoded \textit{channelname} string. The \textit{channelname} string is immediately followed by the payload. Large messages are fragmented into multiple smaller transportation units to achieve a maximum message size of 4 GB, in this case a slightly more complicated header is used, but omitted here.

\section{Attacker Model and Security Goals}
\label{sec:Attacker Model and Security Goals} 

We consider active and modifying attackers in the system. Security is provided only against
outsiders: we do not consider an attacker who has the permission to send on
the multicast group in question (please refer to the discussion on permission management in
Section \ref{ssub:Certificate and Permission Management}). The attacker has
considerable, but limited resources and cannot break common cryptographic
primitives.

Since \textit{channelnames} in LCM are usually domain-specific topics, they should
remain confidential. LCMsec aims to provide confidentiality and integrity of
not only the messages in transit, but also the \textit{channelname} associated with
them. We also provide a notion of authenticity: messages are guaranteed to have
originated from a trusted entity within the LCMDomain, but cannot be attributed
to a specific entity.

We provide a reduced form of security against an attacker, who has no
permission to send on a specific channel, but can send on some other channel
within the group. Against this type of attacker, the integrity and
accountability guarantees remain unchanged, however, confidentiality is provided
only for the contents of messages, not for the channelname (or topic)
associated with messages. We elaborate on the reason for this trade-off in
Section \ref{ssub:Security of messages in transit}.

\section {LCMsec: The Proposed Protocol}
\label {sec: protocol_description}

This section describes the LCMsec protocol in detail. LCMSec employs a
hybrid cryptographic system: Messages in transit are encrypted and authenticated using
symmetric-key cryptography to achieve confidentiality, integrity and
authenticity as outlined in Section \ref{sec:Attacker Model and Security
Goals}. The symmetric key used to this end is generated by an authenticated
group key agreement protocol that does not depend on any central instance to
facilitate. We assume however that each participant possesses a digital
identity with which he can express his rights to the system, details on this
can be found in Section \ref{ssub:Certificate and Permission Management}.

In the following, we first present our solution for securing the messages under the assumption that each participant already has knowledge of required keying material. The generation of this keying material is discussed subsequently.

\subsection{Security of Messages in Transit}%
\label{ssub:Security of messages in transit}

\begin{figure}
    \begin{center}
        \includegraphics[scale=0.6]{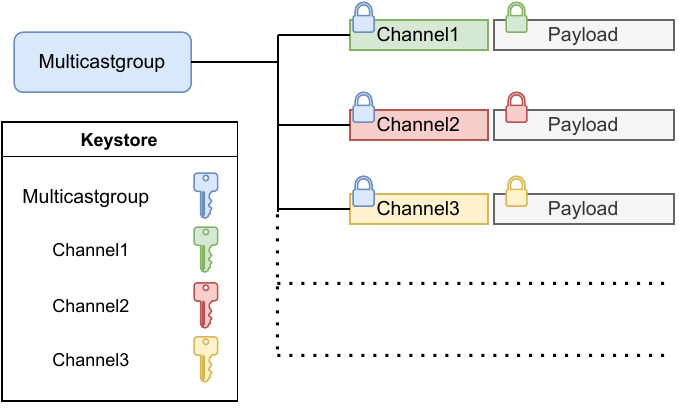}
    \end{center}
    \caption{Hierarchical encryption of channelname and payload in LCMsec}
    \label{fig:hirarch_encr}
\end{figure}

We maintain the hierarchy between channel and group that is inherent to LCM --
one participant can be active on any number of multicast groups, and on any
number of channels within that group. However, participants should only be able
to read and send messages on the LCMDomains that they have permissions to use. 
Thus, to maintain confidentiality and accountability on a per-channel level, we
use a hierarchical scheme illustrated in Figure \ref{fig:hirarch_encr}: one
key, $k_g$, is used to secure the \textit{channelname}. This key is shared between all
users with the permission to access the multicast group. A second key, 
$k_{ch}$, is used to secure the message itself --- this key is shared by all
users with permission to access the LCMDomain.

A receiver can use $k_g$ to decrypt the \textit{channelname}, then look
up the associated $k_{ch}$ to decrypt the message. This carries with it a concession in terms of confidentiality: If an attacker
has access to $k_g$ (he might have access to another channel within that
group), he can learn the \textit{channelname} of messages on other channels. However,
the alternative -- encrypting the \textit{channelname} and payload with a single key
which is unique to the LCMDomain -- would require a subscriber to attempt
decryption of the message with every key that he knows for the group, until he
succeeds. This clearly does not scale for many topics in one group.

\subsubsection{Symmetric encryption of LCM messages}
\label{ssub:lcmsymmetric} We ensure confidentiality and authenticity of LCMsec messages through the use
of authenticated encryption. Specifically, we use AES in Galois/Counter Mode
(GCM) in accordance with the NIST recommendations
\cite{dworkinRecommendationBlockCipher2007}. Using the GCM mode of operation
requires specifying an Initialisation Vector (IV), which must be unique for
each message encrypted with the same key. 

\begin{figure}[!htb]
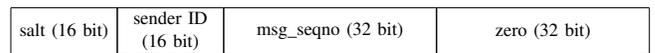

    \begin{bytefield}[bitheight=2.5em]{8} \scalebox{0.7}{
\noindent\makebox[\textwidth-165pt]{%
    \bitbox{9}{salt (16 bit)} & \bitbox{9}{sender ID\\(16 bit)} & \bitbox{18}{msg\_seqno (32 bit)} & \bitbox{18}{zero (32 bit)} 
}
}

\end{bytefield}
\caption{Illustration of the IV used to encrypt LCMsec messages}
\label{fig:IV_illustration}
\end{figure}

While a sequence number is already part of the LCM header, multiple parties
might be communicating on one channel with the same key. Since they increment
their sequence number separately, we also need to uniquely identify senders to
form a unique IV. To this end, we use a 16-bit sender ID. According to the NIST
recommendations, we construct a deterministic 96-bit IV as shown in Figure
\ref{fig:IV_illustration}. The salt, which has not yet been discussed, will be generated as
part of the keying material described in Section \ref{sub:Group Discovery}.

\begin{figure}[!htb] \centering
    \includegraphics[scale=0.8]{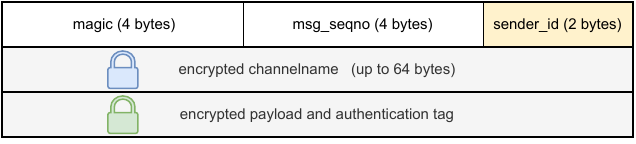} \caption{LCMsec
    Packet format} \label{fig:packet_format}
\end{figure}

The LCMsec packet format shown in Figure \ref{fig:packet_format} is similar to
the LCM packet format. The fields are explained in the following:

\begin{description}
    \item[magic]: Number used to identify LCMsec protocol messages.
    \item[msg\_seqno]: Message sequence number.
    \item[sender\_id]: Unique identifier associated with the node sending the message.
        Its generation is covered in section \ref{sub:Group Discovery and Key Agreement}.
    \item[channelname]: Zero-terminated and ASCII-encoded \textit{channelname}, encrypted
        with $k_{g}$ and AES-CTR. A receiver can decrypt the channelnname bytewise until finding the null-terminator.
        Unauthenticated encryption is used for the \textit{channelname} in
        order save the overhead of a separate authentication tag.
        Authentication of the \textit{channelname} is instead guaranteed by including it in
        the authentication of the payload.
    \item[payload]: The AES/GCM encrypted message including authentication
        tag, encrypted with $k_{ch}$. The \textit{channelname} is included as associated data of the AES/GCM mode.
\end{description}

The spatial overhead of the scheme compared to LCM is thus 18 bytes: two for the $sender\_id$ and 16 for the authentication tag produced by GCM.

\subsubsection{Fragmented messages}%
\label{ssub:Fragmented Messages} As explained in Section \ref{sec:Description
of LCM}, messages that do not fit into a single UDP  packet are supported by
LCM and called fragmented messages. In LCMsec, we encrypt and authenticate
these messages before they are fragmented (and conversely decrypt and verify
them after they are joined back together). This approach authenticates not only the
content of the fragment, but also their order.

\subsubsection{Out-of-order messages and replay attacks}  \label{ssub:Out-of-order Messages} 

Since LCM employs UDP-multicast, messages might arrive out of order. However,
with the introduction of sender IDs, the pair $msg_{id}=(sender\_id, seqno)$ is a
unique identifier for each message. Therefore, it now becomes feasible to
detect and discard or even correct the order of out-of-order messages. Such
behaviour may optionally be configured in LCMsec.

More importantly, the $msg_{id}$ functions to prevent replay attacks. To keep
track of already received messages, a sliding window of the greatest sequence
number received for each peer can be used, in addition to a window of
previously received messages. To efficiently keep track of this
window, the algorithm in appendix C of RFC2401
\cite{atkinsonSecurityArchitectureInternet1998a} or RFC6479
\cite{zouIPsecAntiReplayAlgorithm2012} can be used.

\subsection{Group Discovery and Key Agreement} \label{sub:Group Discovery and Key Agreement}

This section describe how the shared symmetric keying material is generated.
Sharing a key with other users is only meaningful if a notion of identity and
associated permissions exists -- specifically, the permission to send or receive on the LCMDomain. The scheme used to this end is described in Section \ref{ssub:Certificate and Permission Management}.

Subsequently, we will describe the protocol used to perform the key agreement
on the group. We use the Dutta Barua group key agreement \cite{duttaProvablySecureConstant2008} to generate a key among
participants, but this does not suffice: it is simply the backing algorithm
used to perform the key agreement. Thus, the key agreement
process is split into two phases. The first one is a setup phase, which aims to achieve
consensus within the group on the parameters used to perform the DBGKA -- we will call this phase group discovery, described in Section
\ref{ssub:Group Discovery}. The second one is the DBGKA itself
which establishes the shared group key, to be described in Section
\ref{ssub:Dutta Barua Key Agreement}.

Here, we only discuss the key agreement protocol for a
single LCMDomain. In the case of multiple channels, multiple runs of this
protocol will be performed. Indeed, most of the time, at least two
runs of the key agreement protocol will be performed simultaneously: one to
generate $k_g$, another one to generate $k_{ch}$.

\subsubsection{Certificate and permission management}%
\label{ssub:Certificate and Permission Management}

Certificate and permission management is not the main focus of this work, and the solution presented here can easily be changed or extended: it is not tightly coupled to the other areas of this work. Nevertheless, we present an attribute-based access control mechanism based on X.509 certificates \cite{boeyenInternet509Public2008} that is used to both identify participants and manage their permissions.

A user $U$ has access to a specific LCMDomain $L$ if it possesses a valid
X.509 certificate which includes an identifier $\text{ID}_U$ that uniquely
identifies it on the LCMDomain $L$. This $\text{ID}_U$, which is understood to
be the identity for that user on $L$ is encoded into the URN of the Subject
Alternative Name Extension (SAN) of the certificate in accordance with RFC 5280
\cite{boeyenInternet509Public2008}. 
A Certificate
Authority (CA) can issue this certificate and generate the unique identifier
for each domain by incrementing it. The SAN's used shall be of the form

\begin{lstlisting}
urn:lcmsec:<group>:<channel>:<id>
\end{lstlisting}

Multiple SANs can be present in one certificate, enabling an entity to be active on multiple LCMDomains.

\subsubsection{Dutta Barua key agreement}%
\label{ssub:Dutta Barua Key Agreement}

To agree on a key among entities of an LCMDomain, the Dutta Barua authenticated
group key agreement (DBGKA) is used
\cite{duttaProvablySecureConstant2008}\footnote{Two attacks on the DBGKA protocol
have been presented in \cite{zhangTwoAttacksDutta2012}. We have analysed the attacks and conclude that they are not relevant for our solution. Details on this are given in Appendix \ref{sec:attacks}.}.  The protocol
is Diffie-Hellman based and has a number of properties that are interesting
for our use-case. Namely, it is a two-round key-agreement algorithm that
uses broadcast in the second round, which fits the communication topology
used in LCM. Additionally, it is a dynamic protocol: Entities can
\textit{join} a group of users that have already agreed upon a key amongst
themselves while taking advantage of previously computed values, greatly
increasing scalability by reducing both the number of network
transmissions and computations that need to be performed.

The DBGKA provides three operations:

\begin{description}
\item[KeyAgree()]: It allows a number of users to agree on a shared key
\item[Join()]: If a set of users (participants) $P$ has already performed a KeyAgree()
    operation, this operation provides a way for another set of users (joining users), $J$, to agree on
    a shared symmetric key among $P \cup J$. This operation is far more
    efficient than performing KeyAgree($P \cup J$) in terms of network
    usage: in addition to $J$, only 3 users within $P$ need to be active on the network.
\item[Leave()]: Users can leave group, which causes a new key to be
    generated among the remaining users.
\end{description}

However, to use the DBGKA in practice, we need an additional phase which serves
to (1) discover peers and arranges them in a circle, (2) exchange the
certificates of each participant and (3) synchronise the start of the key
agreement operations. In the brokerless spirit of LCM, we aim to achieve these
prerequisites \textit{without} a central instance to coordinate. We will call
the protocol we use to achieve this the LCMsec group discovery protocol, to be
presented in the following section.

\subsubsection{Group discovery}
\label{ssub:Group Discovery}

As discussed in Section \ref{ssub:Certificate and Permission Management}, a group $G$ of entities might have a certificate that grants them permission to be active on an LCMDomain. However, only a subset of these may be active at a specific point in time -- e.g., certain devices may be turned off or disconnected from the system in an IoT context. Within $G$, we define two subsets: Firstly $P$ denotes the set of entities that have already agreed upon a shared secret. Secondly, $J$ consists of the entities that are connected to the network and have expressed their intention to join $P$.

\begin{figure}[!htb] \centering
    \scalebox {0.85}{

\definecolor{cyan}{HTML}{DAE8FC}
\definecolor{green}{HTML}{D5E8D4 }
\definecolor{yellow}{HTML}{FFF2CC }

\begin{tikzpicture}
\begin{scope} [fill opacity = 1.0]
    \draw[fill=cyan, draw = black] (0,0) ellipse (3cm and 2cm);
    \draw[fill=green, draw = black] (-1.5,0) circle (1.3);
    \draw[fill=yellow, draw = black] (1.5,0) circle (1.3);
    \node at (0,1.2) {{$G$}};
    \node at (-1.5,0) {{$P$ Participants}};
    \node at (1.5,0) {{$J$ Joining}};
    \end{scope}
\end{tikzpicture}
    }
    \caption{Entities on the LCMDomain}
    \label{fig:LCMDomain_Venn}
\end{figure}
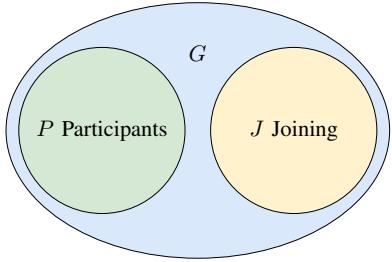

First, we note that for the purposes of the group discovery protocol, there is no need for a separate initial KeyAgree() and subsequent Join() operation: without loss of generality, $P$ may be empty, and both cases can be handled by a Join() operation.

Additionally, we note that the problem of arranging participants in a circle is
equivalent to achieving consensus on the sets of $P$ and $J$ among all $U \in P
\cup J$. They can then be ordered by their unique identifiers ($ID_U$).
Alternatively, a hash of their certificate could be used. Subsequently, a
deterministic mapping to sender IDs (that is, an unsigned integer which fits
into 16 bits) can be performed. The synchronisation of the KeyAgree() operation
can also be regarded as part of this consensus problem: the consensus on a
timestamp $t$ at which the key agreement shall commence.

The problem of consensus in distributed computing is well-studied. A popular
solution is the RAFT Protocol \cite{ongaroSearchUnderstandableConsensus2014},
which achieves consensus among a group of distributed nodes by voting for a
leader via a randomised timeout, who then replicates a log data structure to all
other nodes. In our group discovery protocol, we take the lessons learned from
RAFT and adapt them to our use-case by noticing that replication of a
log data structure is not what we desire: We do not care about consensus on data
in the past, only the current sets $P$ and $J$ are of interest. Additionally,
we do not require a strict form of consensus: The DBGKA will reliably
fail if there is no consensus on the participants involved (instead of
producing an invalid key). Finally, we notice that RAFT uses \textit{heartbeats} to
ensure that a leader always exists, which is problematic in a multicast
communication topology due to scalability issues. However, a leader
is not always needed, but only when a Join() operation is initiated.

We thus present the central idea of our group discovery protocol. Unlike RAFT,
we form consensus only on an as-needed basis (that is, whenever a new key is
necessary) and vote for a leader not via timeout, but instead form consensus
\textit{on the data itself}. By defining $(P, J, t)
\in \mathcal{D}$, we can impose a weak order on $\mathcal{D}$: for  $D_1 =
(P_1, J_1, t_1)$  and $D_2 = (P_2, J_2, t_2)$, $D1, D2 \in \mathcal{D}$,
\begin{equation}
    \begin{split}
        D_1 \leq D_2 \iff &(|P_1| \leq |P_2|) \vee \\
                          &( |P_1| = |P_2| \wedge |J_1| \leq |J_2|) \vee \\
                          &(|P_1| = |P_2| \wedge |J_1| = |J_2| \wedge t_1 >=
                          t_2)
    \end{split}
\end{equation}

\begin{figure}[!htb]\centering
    \centering\includegraphics[scale=0.45]{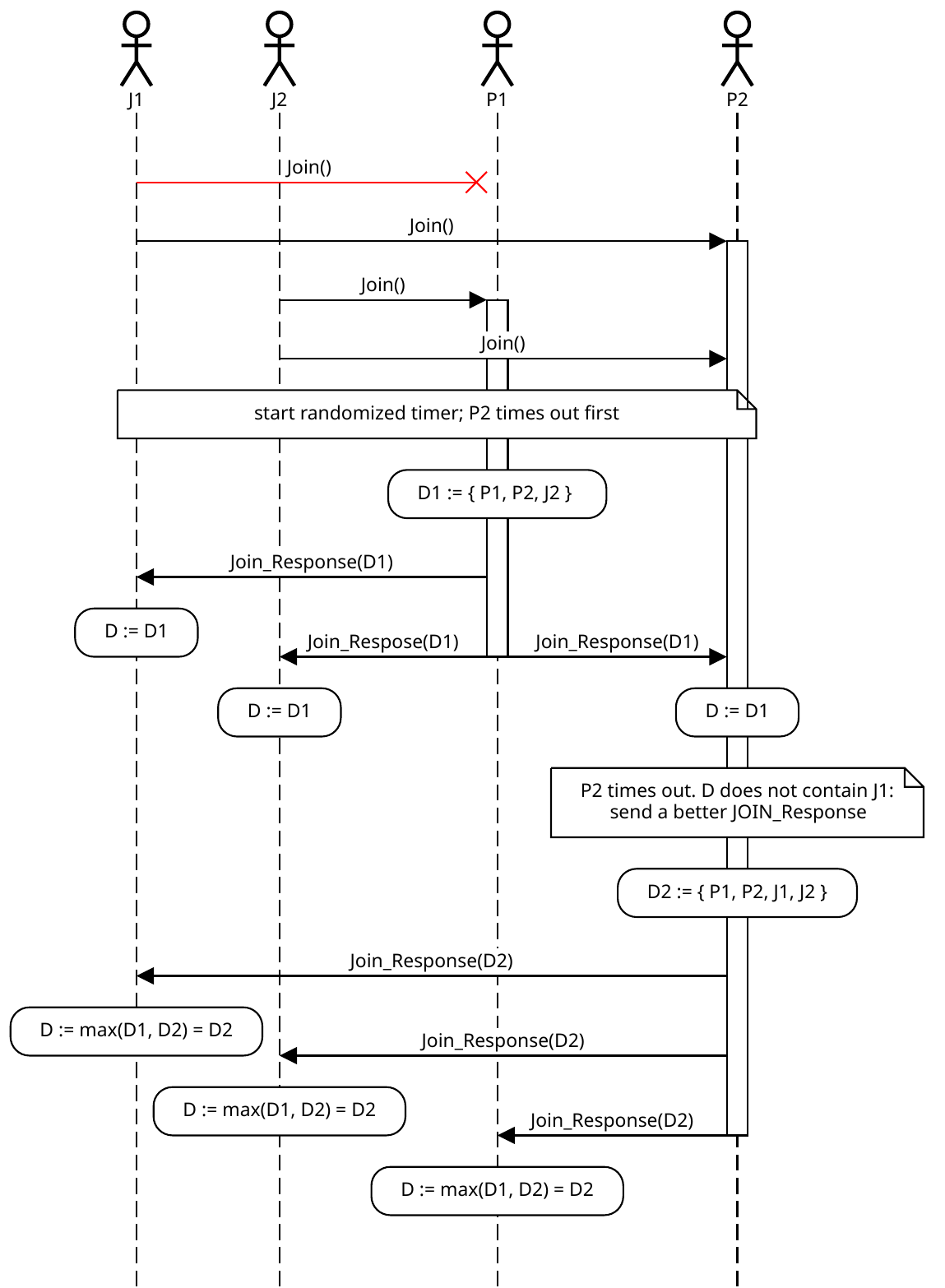} 
    \caption{Sequence diagram illustrating a simplified version of the LCMsec
        Group Discovery Protocol in the case of a lost message. A similar
        situation arises if the Join() is delayed instead of lost. Additionally,
    Join() messages between J1 and J2 exist, but are omitted for brevity.}
    \label{fig:groupconsensus_2}
\end{figure}

By adding a small, random offset $\varepsilon$ to $t$, this weak order can be transformed
into a total order. The way we define this order is not arbitrary: we maximise
$|J|$ and $|P|$, while minimising $t$ to guarantee termination of the discovery
phase. Consensus is now simply achieved by each participant keeping track of the largest $D$ it has
observed.

With these considerations in mind, we will now describe the  discovery protocol
in detail. All messages are authenticated with a DSA, but the signatures -- as
well as the verification of the signatures (and associated certificates) are
omitted here for brevity. Naturally, LCM is used as a communication medium.

An entity $a \in G$ with a certificate $cert_a$ expresses the intent to
initiate the group discovery and subsequent key agreement on an LCMDomain $L$ by
transmitting $JOIN_a = (t_a, cert_a)$ with $t_a =
t_{now}+\varepsilon$ on $L$. Additionally, it initialises $D_a := (\emptyset,
\left\{ cert_a \right\}, t_a)$.

Upon receiving such a $JOIN$, entity $b$ with $D_b = (P_b, J_b, t_b)$, stores the
certificate contained for subsequent use. After a randomised delay, a number of
such JOINs may have been received -- we will call the set comprising them
$M$. The set $J_{new} = M \setminus J_b$ then describes the JOINs that have been observed by $b$, but are not yet answered.
If $J_{new} \neq \emptyset$, $b$ now sets $J_b := J_b \cup J_{new}$ and $t_b := \min(t_b, \min (t \mid (t, cert) \in J_{new}) $ before transmitting $JOIN\_Response = D_b$.

Any entity $c$ with $D_c$, upon receiving $JOIN\_Response=D_r$ stores the
included certificates for later use and sets $D_c:=\max(D_c, D_r)$.

Once $t$ has been reached, the DBGKA will be initiated and  no further
modification to $D$ is permitted until it fails or succeeds.
If successful, each participant will set $P:= P \cup J$ and $J :=
\emptyset$, otherwise they will set  $P:=\emptyset$ and $J:=
\emptyset$ and restart the group discovery phase by transmitting a $JOIN$.

\section{Implementation and Evaluation}%
\label{sec:Implementation}

An implementation of LCMsec is publicly available\footnote{\url{https://github.com/Barkhausen-Institut/lcm-sec}}. It is written in
C++ and uses the Botan\footnote{\url{https://botan.randombit.net/}} cryptography library. In the implementation of the
Dutta Barua protocol, we use a modified version based on elliptic curve
cryptography for performance reasons.

\subsection{Latency and Throughput}
\label{ssub:Latency and throughput}

\begin{figure}[!htb] \centering
    \includegraphics[scale=0.58]{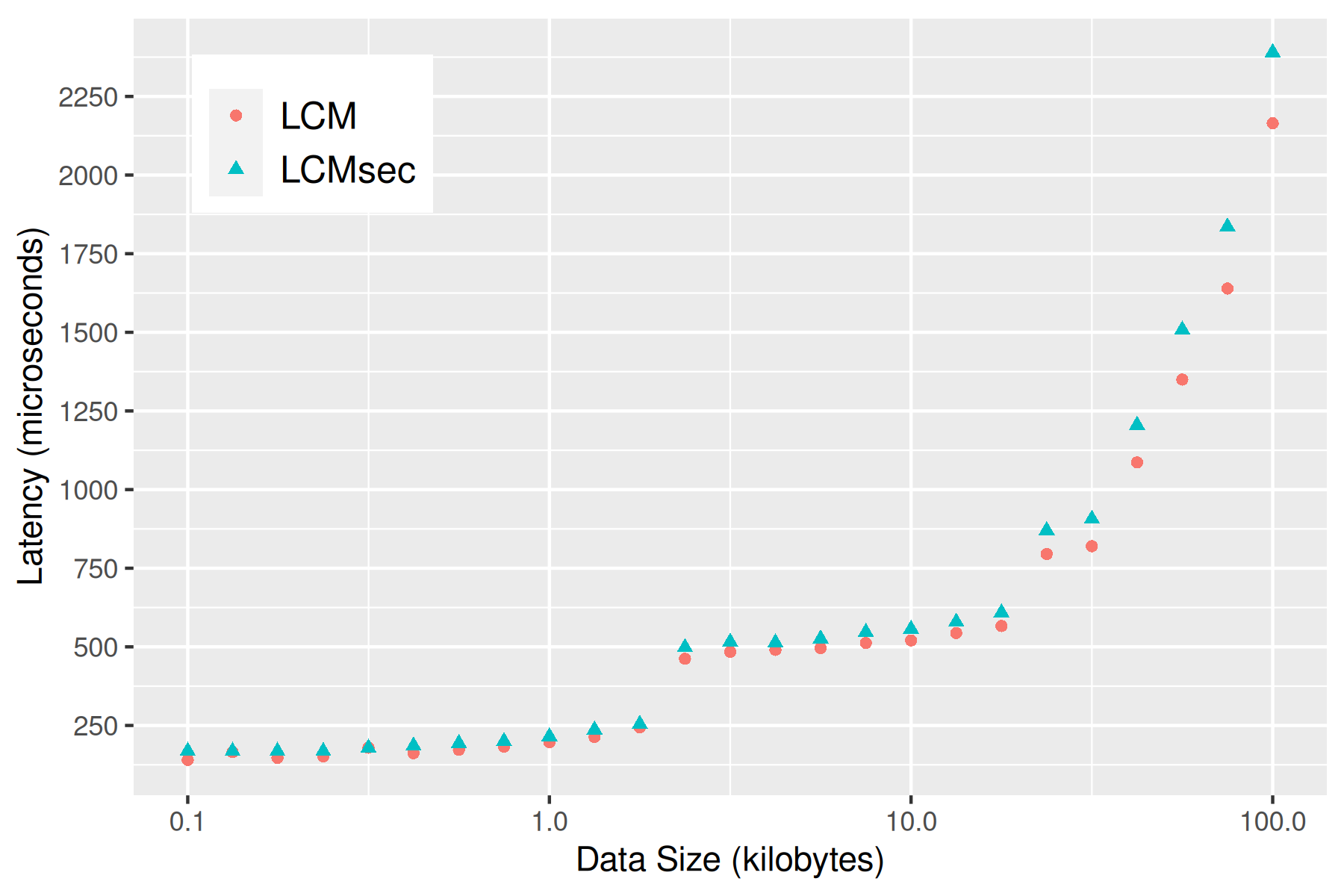}
    \caption{Latency comparison between LCM and LCMsec }
    \label{fig:latency}
\end{figure}

Latency and throughput of the LCMsec protocol were tested using two identical servers with an Intel Xeon Gold 5317 processor and 8GB RAM
running Linux 5.15. The servers were one hop apart with a 1GBit/s link between them. 
To test the latency of LCMsec messages, an echo test was performed: one of the servers, the source, transmitted
messages of sizes ranging from 100 Bytes to 100 Kilobytes. Upon receiving one of
these messages, the other server immediately re-transmitted it. Upon receiving
the original message back, the latency was measured by the source. For each message size, a
total of 1000 latency measurements were taken. The same was done for the original LCM
library. The results are depicted in Figure \ref{fig:latency} -- as one can
see, there is only a small latency overhead. Note that the jump at 3 KB is due
to fragmentation of the LCM messages, which occurs at that size.

To measure the throughput achieved by LCMsec, a similar echo test was performed
on the same servers. Using a fixed message size, the source increased the bandwidth at which it transmitted while recording the
number of messages it received back. In such a test, the percentage of lost messages can
indicate the throughput capabilities of LCMsec. However, no
difference between LCM and LCMsec was observed: in both cases, no messages were lost up to a
bandwidth of 123MB/s. After this point, a majority of messages were dropped
since the limit of the link between the servers had been reached.

\subsection{Evaluation of the Group Discovery}%
\label{sub:Group Discovery}

The most expensive part of the group discovery are the JOIN\_Responses: They
may be large since they contain the certificates of all other users. Thus, the
number of JOIN\_Responses needed should be kept to a minimum. To evaluate the
performance of the protocol, measuring the time taken to perform the group
discovery protocol is not helpful, since it is bounded by timeouts. Instead, we
count the number of JOINs and JOIN\_Responses transmitted while a varying
number of nodes  execute the group discovery protocol and subsequent DBGKA
twice (in order to agree on both $k_g$ and $k_{ch}$).

\begin{figure}[!htb] \centering
    \includegraphics[scale=0.55]{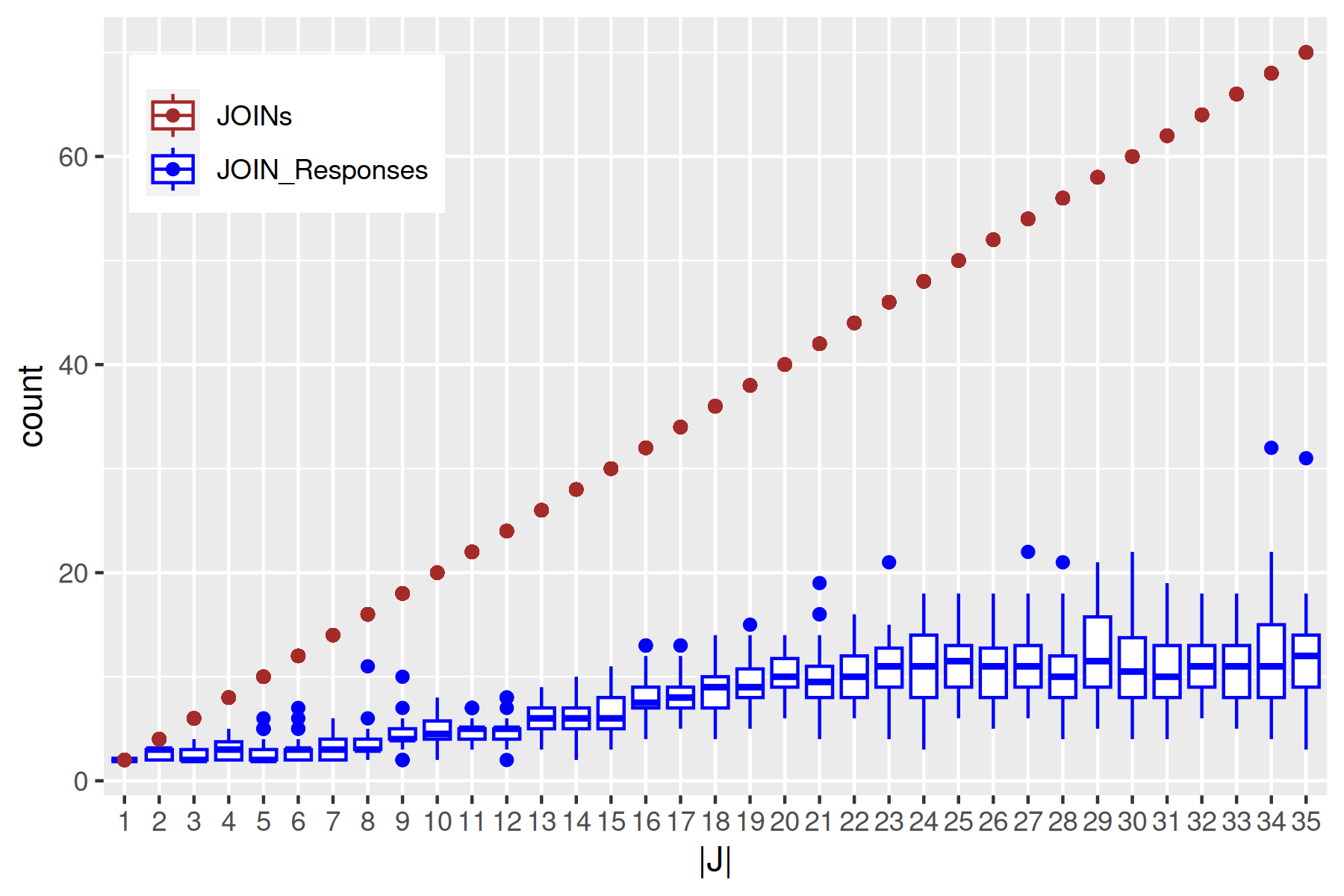}
    \caption{Performing the group discovery and key agreement protocol with $|P|=0$, varying $|J|$ and emulated network delays}
    \label{fig:consensus_intial_delay}
\end{figure}

Additionally, the Linux NetEm facility was
used to emulate noramlly distributed ($\mu = 25ms$, $\sigma^2 = 5ms$)
network delays, affecting all messages used during the consensus and key
agreement. The results are shown in Figure \ref{fig:consensus_intial_delay}.
While the chosen distribution is somewhat arbitrary, the results show not only
that the group consensus protocol performs in real-life networks with a large
number of participants, but also a certain resilience of the consensus
protocol.

\section{Conclusion}

In this work, we presented LCMsec, a new secure brokerless Publish/Subscribe
protocol based on UDP multicast. We have added confidentiality, integrity and
authenticity to the existing LCM protocol while minimising both overhead and
computational complexity. LCMsec can be used in most environments in
which LCM is currently used, e.g., IoT, automotive and robotics applications. This has been achieved by using a different threat model
than previous work in the domain of multicast authentication. We make no
distinction between subscribers and publishers, each subscriber is also allowed
to publish messages. However, an attribute-based access
control mechanism is available through the use X.509 certificates that grants
access only to specific LCMdomains.

LCMsec is decentralised in the sense that there is no need for a central server
to broker messages, facilitate key exchanges or discover peers. A discovery
mechanism is instead built-in, which facilitates ease-of-use and flexibility.
Despite the shared symmetric key, it should be noted that the protocol is
scalable in dynamic situations: Through use of the Dutta-Barua group key
agreement, the number of network interactions when a publisher or subscriber
joins a topic is minimised.

\appendix

\subsection{Two attacks on the Dutta-Barua group key agreement}
\label{sec:attacks}

Zhang et al. present two attacks on the DBGKA
protocol \cite{zhangTwoAttacksDutta2012}. To fully understand them and this section, some understanding of the
Dutta-Barua protocol \cite{duttaProvablySecureConstant2008} is required. While
a full review of the protocol is out of scope for this work, for the purposes
of this section, the most important thing is to understand that each KeyAgree()
and Join() operation is associated with an \textbf{instance id} $d$. This
instance id is incremented for each of those operations and can never be reused.
Note that $d$ can be regarded as a nonce: while it is not random, it is never
reused. Another example of a protocol that uses non-random nonces is Wireguard
\cite{donenfeldWireGuardNextGeneration2017}.

Both attacks described by Zhang et al. are carried out by one or
multiple malicious users who are part of the Dutta-Barua group, that is they
have successfully participated in the Dutta-Barua key agreement in the past. In
this sense, the premise of the DBGKA is already violated: The DBGKA protocol
provides no security against malicious insiders. Nevertheless, one should take this
form of attack seriously: An honest user - representing, for instance, an IoT
device - might at some point be compromised and \textit{become} dishonest.
Alternatively, he might have been dishonest all along, but his certificate is
only revoked at a later stage. We will therefore discuss both attacks and show
why they pose no threat to the LCMsec protocol.

\subsubsection{First Attack}

The first attack is carried out by a malicious leaving user who has been part
of a previous successful Dutta-Barua \textit{KeyAgreement()} operation during which he has made
some preparation for the attack by storing some of the protocol messages. When
the \textit{Leave()} operation is executed to expel this user from the group, Zhang et.
al. show that the attacker can compute the new session key using the values he stored
earlier.

However, as we understand the DBGKA, the purpose of the \textit{Leave()}
operation is not to expel dishonest users, but as a way for honest users to
leave. When an honest user leaves in this way, it is possible for the remaining
users to efficiently agree on a new key. If an honest user, on the other hand,
does not execute the \textit{Leave()} operation, a new \textit{KeyAgreement()} operation
has to performed, which is a lot less efficient for large groups. To expel a
malicious user, the remaining users instead execute the \textit{KeyAgree()} operation
amongst themselves -- this way, the attack is bypassed entirely.

Note that in the current version of LCMsec, we do not include a mechanism for
certificate revocation or expelling users from the group and make no use of the
\textit{Leave()} operation, so this attack does not concern us. Still, the ability to
add such a feature in the future is important. As we discussed,
this can be done safely by using the \textit{KeyAgree()} operation whenever a
certificate is revoked.

\subsubsection{Second Attack}

The second attack is a replay attack that is carried out by two cooperating,
malicious users $U_i$ and $U_j$ that have been part of a Dutta-Barua key agreement. For
simplicity and without loss of generality, we assume here that for this first
KeyAgree() operation, the associated instance number of all users during this
was $d=1$. By storing some of the messages during the second round
of the protocol, the authors claim that $U_i$ and $U_j$ with $j>i+1$ are
able to impersonate all the users $U_k$, $i < k < j$ between them (with respect
to the circle on which users are arranged) during a subsequent \textit{Join()}
operation. The authors claim that this attack is possible since the DB-Protocol
does not use nonces, which is the mechanism they say it should to prevent this
attack.

However, as discussed earlier, the instance id $d$ is a nonce, though it is not
a random one. Note that the round-2 messages of the DBGKA are of the form
$M_k=(U_k|2|Y_k|d_k)$, where $U_k=k$ is the id of the user $U_k$, $2$ indicates that
it is the message for the \textit{second} round of the protocol, $Y_k$ is the
result of the computation for that round and user $k$, and $d_k=1$ is the
instance id of user $k$. Note also that the \textit{transmitted} message during
the second round is $M|\sigma_k$, where $\sigma_k$ is a signature over $M_k$
computed with the private key known only by user $k$. The malicious users
$U_j$ and $U_k$ can therefore not modify $M_k|\sigma_k$, they can only store
and replay it.

The actual attack consists of $U_i$ and $U_j$ impersonating $U_k$ by
transmitting the stored round-2 messages $M_k|\sigma_k$ with $d_k=1$. However,
$d_k=1$ has already been used for user $U_k$. Legitimate users will have
observed this during the initial \textit{KeyAgree()} operation and therefore
simply ignore the replayed messages -- the attack fails.

\apptocmd{\sloppy}{\hbadness 10000\relax}{}{}
\bibliographystyle{myIEEEtran}
\bibliography{lcmsec.bib} 

\begin{thebibliography}{10}
\providecommand{\url}[1]{#1}
\csname url@samestyle\endcsname
\providecommand{\newblock}{\relax}
\providecommand{\bibinfo}[2]{#2}
\providecommand{\BIBentrySTDinterwordspacing}{\spaceskip=0pt\relax}
\providecommand{\BIBentryALTinterwordstretchfactor}{4}
\providecommand{\BIBentryALTinterwordspacing}{\spaceskip=\fontdimen2\font plus
\BIBentryALTinterwordstretchfactor\fontdimen3\font minus \fontdimen4\font\relax}
\providecommand{\BIBforeignlanguage}[2]{{%
\expandafter\ifx\csname l@#1\endcsname\relax
\typeout{** WARNING: IEEEtran.bst: No hyphenation pattern has been}%
\typeout{** loaded for the language `#1'. Using the pattern for}%
\typeout{** the default language instead.}%
\else
\language=\csname l@#1\endcsname
\fi
#2}}
\providecommand{\BIBdecl}{\relax}
\BIBdecl

\bibitem{eugsterManyFacesPublish2003}
\BIBentryALTinterwordspacing
P.~T. Eugster, P.~A. Felber, R.~Guerraoui, and A.-M. Kermarrec, ``The many faces of publish/subscribe,'' \emph{ACM Computing Surveys}, vol.~35, no.~2, pp. 114--131, Jun. 2003.
\BIBentrySTDinterwordspacing

\bibitem{huangLCMLightweightCommunications2010}
\BIBentryALTinterwordspacing
A.~S. Huang, E.~Olson, and D.~C. Moore, ``{{LCM}}: {{Lightweight Communications}} and {{Marshalling}},'' in \emph{2010 {{IEEE}}/{{RSJ International Conference}} on {{Intelligent Robots}} and {{Systems}}}, pp. 4057--4062.
\BIBentrySTDinterwordspacing

\bibitem{canettiMulticastSecurityTaxonomy1999}
\BIBentryALTinterwordspacing
R.~Canetti, J.~Garay, G.~Itkis, D.~Micciancio, M.~Naor, and B.~Pinkas, ``Multicast security: A taxonomy and some efficient constructions,'' in \emph{{{IEEE INFOCOM}} '99. {{Conference}} on {{Computer Communications}}. {{Proceedings}}. {{Eighteenth Annual Joint Conference}} of the {{IEEE Computer}} and {{Communications Societies}}. {{The Future}} Is {{Now}} ({{Cat}}. {{No}}.{{99CH36320}})}, pp. 708--716 vol.2.
\BIBentrySTDinterwordspacing

\bibitem{MessageQueuingTelemetry}
\BIBentryALTinterwordspacing
``Message {{Queuing Telemetry Transport}}.'' [Online]. Available: \url{https://mqtt.org/}
\BIBentrySTDinterwordspacing

\bibitem{pardo-castelloteOMGDataDistributionService2003}
G.~{Pardo-Castellote}, ``{{OMG Data-Distribution Service}}: Architectural overview,'' in \emph{23rd {{International Conference}} on {{Distributed Computing Systems Workshops}}, 2003. {{Proceedings}}.}, May 2003, pp. 200--206.

\bibitem{deeringHostExtensionsIP1989}
\BIBentryALTinterwordspacing
S.~E. Deering, ``Host extensions for {{IP}} multicasting,'' {Internet Engineering Task Force}, Request for {{Comments}} RFC 1112, Aug. 1989.
\BIBentrySTDinterwordspacing

\bibitem{onicaScalableDependablePrivacyPreserving2016}
\BIBentryALTinterwordspacing
E.~Onica, P.~Felber, H.~Mercier, and E.~Rivi{\`e}re, ``Towards {{Scalable}} and {{Dependable Privacy-Preserving Publish}}/{{Subscribe Services}},'' in \emph{Fast {{Abstract}} in the 46th {{Annual IEEE}}/{{IFIP International Conference}} on {{Dependable Systems}} and {{Networks}}}, M.~Roy, J.~A. Lopez, and A.~Casimiro, Eds., {Toulouse, France}, Jun. 2016.
\BIBentrySTDinterwordspacing

\bibitem{bernardFrameworkSecurePrivate2010}
S.~Bernard, M.~G. {Potop-Butucaru}, and S.~Tixeuil, ``A {{Framework}} for {{Secure}} and {{Private P2P Publish}}/{{Subscribe}},'' in \emph{Stabilization, {{Safety}}, and {{Security}} of {{Distributed Systems}}}, ser. Lecture {{Notes}} in {{Computer Science}}, S.~Dolev, J.~Cobb, M.~Fischer, and M.~Yung, Eds., pp. 531--545.

\bibitem{malinaSecurePublishSubscribe2019}
\BIBentryALTinterwordspacing
L.~Malina, G.~Srivastava, P.~Dzurenda, J.~Hajny, and R.~Fujdiak, ``A {{Secure Publish}}/{{Subscribe Protocol}} for {{Internet}} of {{Things}},'' in \emph{Proceedings of the 14th {{International Conference}} on {{Availability}}, {{Reliability}} and {{Security}}}, ser. {{ARES}} '19, pp. 1--10.
\BIBentrySTDinterwordspacing

\bibitem{ionSupportingPublicationSubscription2010}
M.~Ion, G.~Russello, and B.~Crispo, ``Supporting {{Publication}} and {{Subscription Confidentiality}} in {{Pub}}/{{Sub Networks}},'' in \emph{Security and {{Privacy}} in {{Communication Networks}}}, ser. Lecture {{Notes}} of the {{Institute}} for {{Computer Sciences}}, {{Social Informatics}} and {{Telecommunications Engineering}}, S.~Jajodia and J.~Zhou, Eds., pp. 272--289.

\bibitem{hamadSPPSSecurePolicybased2021}
M.~Hamad, E.~Regnath, J.~Lauinger, V.~Prevelakis, and S.~Steinhorst, ``{{SPPS}}: {{Secure Policy-based Publish}}/{{Subscribe System}} for {{V2C Communication}},'' in \emph{2021 {{Design}}, {{Automation}} \& {{Test}} in {{Europe Conference}} \& {{Exhibition}} ({{DATE}})}, Feb. 2021, pp. 529--534.

\bibitem{dahlmannsTransparentEndtoEndSecurity2021}
\BIBentryALTinterwordspacing
M.~Dahlmanns, J.~Pennekamp, I.~B. Fink, B.~Schoolmann, K.~Wehrle, and M.~Henze, ``Transparent {{End-to-End Security}} for {{Publish}}/{{Subscribe Communication}} in {{Cyber-Physical Systems}},'' in \emph{Proceedings of the 2021 {{ACM Workshop}} on {{Secure}} and {{Trustworthy Cyber-Physical Systems}}}, ser. {{SAT-CPS}} '21, pp. 78--87.
\BIBentrySTDinterwordspacing

\bibitem{imatixcorporationZeroMQBrokerVs}
\BIBentryALTinterwordspacing
{iMatix Corporation}, ``{{ZeroMQ Broker}} vs . {{Brokerless Messaging}} - {{Whitepaper}}.'' [Online]. Available: \url{http://wiki.zeromq.org/whitepapers:brokerless}
\BIBentrySTDinterwordspacing

\bibitem{imatixcorporationCurveZMQProtocolSecure2023}
\BIBentryALTinterwordspacing
iMatix Corporation, ``{{CurveZMQ}} - a protocol for secure messaging across the {{Internet}},'' Jan. 2023. [Online]. Available: \url{http://rfc.zeromq.org/spec/26/}
\BIBentrySTDinterwordspacing

\bibitem{objectmanagementgroupDDSSecuritySpecification2018}
\BIBentryALTinterwordspacing
{Object Management Group}, ``{{DDS Security Specification}},'' Jun. 2018. [Online]. Available: \url{https://www.omg.org/spec/DDS-SECURITY/1.1/PDF}
\BIBentrySTDinterwordspacing

\bibitem{kimSecurityPerformanceConsiderations2018}
\BIBentryALTinterwordspacing
J.~Kim, J.~M. Smereka, C.~Cheung, S.~Nepal, and M.~Grobler, ``Security and {{Performance Considerations}} in {{ROS}} 2: {{A Balancing Act}},'' Sep. 2018. [Online]. Available: \url{http://arxiv.org/abs/1809.09566}
\BIBentrySTDinterwordspacing

\bibitem{duttaProvablySecureConstant2008}
R.~Dutta and R.~Barua, ``Provably {{Secure Constant Round Contributory Group Key Agreement}} in {{Dynamic Setting}},'' \emph{IEEE Transactions on Information Theory}, vol.~54, no.~5, pp. 2007--2025, May 2008.

\bibitem{ongaroSearchUnderstandableConsensus2014}
D.~Ongaro and J.~Ousterhout, ``In search of an understandable consensus algorithm,'' in \emph{Proceedings of the 2014 {{USENIX}} Conference on {{USENIX Annual Technical Conference}}}, ser. {{USENIX ATC}}'14, pp. 305--320.

\bibitem{dworkinRecommendationBlockCipher2007}
\BIBentryALTinterwordspacing
M.~J. Dworkin, ``Recommendation for {{Block Cipher Modes}} of {{Operation}}: {{Galois}}/{{Counter Mode}} ({{GCM}}) and {{GMAC}},'' \emph{NIST}, Nov. 2007.
\BIBentrySTDinterwordspacing

\bibitem{atkinsonSecurityArchitectureInternet1998a}
\BIBentryALTinterwordspacing
R.~Atkinson and S.~Kent, ``Security {{Architecture}} for the {{Internet Protocol}},'' {Internet Engineering Task Force}, Request for {{Comments}} RFC 2401, Nov. 1998.
\BIBentrySTDinterwordspacing

\bibitem{zouIPsecAntiReplayAlgorithm2012}
\BIBentryALTinterwordspacing
T.~T.~T. ZOU) and X.~Zhang, ``{{IPsec Anti-Replay Algorithm}} without {{Bit Shifting}},'' {Internet Engineering Task Force}, Request for {{Comments}} RFC 6479, Jan. 2012.
\BIBentrySTDinterwordspacing

\bibitem{boeyenInternet509Public2008}
\BIBentryALTinterwordspacing
S.~Boeyen, S.~Santesson, T.~Polk, R.~Housley, S.~Farrell, and D.~Cooper, ``Internet {{X}}.509 {{Public Key Infrastructure Certificate}} and {{Certificate Revocation List}} ({{CRL}}) {{Profile}},'' {Internet Engineering Task Force}, Request for {{Comments}} RFC 5280, May 2008.
\BIBentrySTDinterwordspacing

\bibitem{zhangTwoAttacksDutta2012}
H.~Zhang, C.~Xu, C.~Li, and A.~R. Sangi, ``Two {{Attacks}} on {{Dutta}}'s {{Dynamic Group Key Agreement Protocol}},'' in \emph{Wireless {{Communications}} and {{Applications}}}, ser. Lecture {{Notes}} of the {{Institute}} for {{Computer Sciences}}, {{Social Informatics}} and {{Telecommunications Engineering}}, P.~S{\'e}nac, M.~Ott, and A.~Seneviratne, Eds., pp. 419--425.

\bibitem{donenfeldWireGuardNextGeneration2017}
\BIBentryALTinterwordspacing
J.~A. Donenfeld, ``{{WireGuard}}: {{Next Generation Kernel Network Tunnel}},'' in \emph{Proceedings 2017 {{Network}} and {{Distributed System Security Symposium}}}.
\BIBentrySTDinterwordspacing

\end{thebibliography}

\end{document}